\documentclass{PoS}

\title{The hidden-charm multiquark states}

\ShortTitle{}

\author{Hua-Xing Chen\\
School of Physics and Beijing Key Laboratory of Advanced Nuclear
Materials and Physics, Beihang University, Beijing 100191, China}
\author{Wei Chen\\
Department of Physics and Engineering Physics, University of
Saskatchewan, Canada}
\author{Jun He\\
Nuclear Theory Group, Institute of Modern Physics of
CAS, Lanzhou 730000, China}
\author{Xiang Liu\\
School of Physical Science and Technology, Lanzhou University,
Lanzhou 730000,  China}
\author{\speaker{Shi-Lin Zhu}\thanks{Plenary talk.}\\
        Department of Physics and State Key Laboratory of Nuclear Physics and Technology \\
and Collaborative Innovation Center of Quantum Matter,\\
Peking University, Beijing 100871, China\\
        E-mail: \email{zhusl@pku.edu.cn}}

\abstract{Since 2003 many charmonium-like states were observed
experimentally. Especially those charged charmonium-like $Z_c$
states and bottomonium-like $Z_b$ states cannot be accommodated
within the naive quark model, which are good candidates of either
the hidden-charm tetraquark states or molecules composed of a pair
of charmed mesons. In 2015, the LHCb Collaboration discovered two
hidden-charm pentaquark states, which are also beyond the quark
model. In this talk, we review the current experimental progress and
investigate various theoretical interpretations of these candidates
of the multiquark states. We list the puzzles and theoretical
challenges of these models when confronted with the experimental
data. We also discuss possible future measurements which may
distinguish the theoretical schemes on the underlying structures of
the hidden-charm multiquark states.}

\FullConference{The 26th International Nuclear Physics Conference\\
        11-16 September, 2016\\
        Adelaide, Australia}

\begin{document}

\section{Motivation}

We all know the motion and interaction of hadrons differ from those
of nuclei and quark or gluons. Hadron physics is the bridge between
nuclear physics and particle physics. The famous Higgs mechanism
contributes around 20 MeV to the nucleon mass through the current
quark mass. Nearly all the mass of the nucleon and visible matter in
our universe comes from the nonperturbative QCD interaction. The
study of hadron spectrum explores the mechanism of confinement and
chiral symmetry breaking ($\chi$SB).

According to the number (N) of the valence quarks, we have the
glueballs (N=0), mesons (N=2), baryons (N=3), tetraquarks (N=4),
pentaquarks (N=5), deuteron (N=6), nuclei, neutron stars. At
present, the tetraquarks and pentaquarks are still missing.

It's interesting to compare QCD and QED. In fact, the QED analogues
of the baryon, glueball and hybrid meson do NOT exist. However,
there are common features between the QED and QCD spectrum. For the
positronium, we have the pion which is composed of $q\bar q$. For
the hydrogen atom, we have the heavy meson and baryon where the
light quarks circle around the heavy quark. For the positronium
molecule, one may expect the light scalar tetraquark candidates such
as the sigma and kappa mesons. For the hydrogen molecule, one may
expect the hidden-charm tetraquark states. For the polarized atoms
or molecules, we have the deuteron. We may also expect other
hadronic molecules composed of heavy hadrons.

The organizers of the conference asked me to tell the audience why
the XYZ states are interesting to nuclear physicists. First of all,
some XYZ state may be shallow deuteron-like states. The chiral
dynamics (or the pion-exchange force) and coupled channel effects
are important. We can use the same nuclear physics techniques to
study some of the XYZ states. In my talk I will mainly focus on the
XYZ states which the audience may have interest in. Interested
readers may also consult the extensive review on the hidden-charm
multiquark states \cite{report}.

\section{Experimental status}

Since 2003, many charmonium-like states were observed. Their
production mechanisms include the initial state radiation, double
charmonium production, two-photon fusion, B meson decay and excited
charmonium decays etc. Their discovery modes include both the
hidden-charm and open-charm modes.

Up to now, the lattice QCD simulation reproduces the charmonium
spectrum below the $D\bar D$ threshold very well. On the other hand,
many new states above the $D\bar D$ threshold were discovered
experimentally since 2003. Some are even charged. They are good
candidates of the exotic mesons.

In the very beginning, I want to emphasize that many XYZ states lie
very close to the open-charm threshold. It's quite possible that
some states are not real resonances. They could be fake signals
arising from
\begin{itemize}
\item  -  Kinematical effect
\item  -  Opening of new threshold
\item  -  Cusp effect
\item  -  Final state interaction
\item  -  Interference between continuum and well-known charmonium states
\item  -  Triangle singularity due to the special kinematics
\item  -  $\cdots$
\end{itemize}

\section{Theory}

Many XYZ states do not fit into quark model spectrum easily. There
are some popular theoretical speculations.
\begin{itemize}
\item  Hadronic molecules are loosely bound states composed
of a pair of heavy hadrons. The long-range pion exchange force may
play an important role in the formation of the loosely bound
hadronic molecules. The molecular states may be quite sensitive to
the isospin configurations.

\item  Tetraquarks are speculated to be tightly bound objects of four quarks.
They are bound by the colored-force between quarks. They may decay
through rearrangement. Since the dominant part of the color
confining force are flavor independent, the tetraquarks shall always
be accompanied by partner states. In general, there are many states
within the same multiplet. Some are even charged or carry
strangeness, which provides a powerful handle in the experimental
search of these states. The color-magnetic interaction is
responsible for the mass splitting between these states. If one
member of the multiplet exists, all the other members should also
exist.

\item  Hybrid charmonium are bound states composed of a pair of
quarks and one or more gluons.

\item  Last but not the least, these XYZ states could also be the
conventional charmonium. One should be very cautious that the quark
model spectrum could be distorted by the coupled-channel effects.

\end{itemize}

\section{Selected examples: $P_c, X(3872), Z_b/Z_c, Y(4260)$}

The multiquark states were first proposed by Gell-Mann in 1964 in
his pioneering paper \cite{Gell-Mann}. However, no convincing states
were discovered in the past several decades. In 2003, LEPS
collaboration reported the $\Theta$ pentaquark \cite{leps}. Now this
signal disappeared.

\subsection{$P_c$ states}

In 2015, the LHCb collaboration reported two hidden-charm pentaquark
states \cite{lhcb}. In the decay process $\Lambda_b\to J/\psi P K$,
LHCb observed two resonances in the $J/\psi P$ final state. The
lower state is broad. Its mass is 4380 MeV and width is around 205
MeV. The higher state lies around 4450 MeV. It's quite narrow. From
the best fit, their spin-parity quantum numbers are ${3\over 2}^-$
and ${5\over 2}^+$ respectively.

According to the color configurations, there are two possible
binding mechanisms: tightly bound or weakly bound. The idea of the
loosely bound molecular states is not new in nuclear physics since
Yukawa proposed the pion in 1935. The deuteron is a very loosely
bound molecular state composed of a proton and neutron arising from
the color-singlet meson exchange.

We adopted the same one-meson-exchange formalism to discuss the
possible molecular states composed of a pair of heavy hadrons. The
charmed meson and baryon are the same as the proton and neutron in
the formation of the loosely bound molecular states. Several years
ago, we studied the hidden-charm molecular baryons composed of
anti-charmed meson and charmed baryon \cite{cpc}. The lower state
$P_c(4380)$ could be explained the ${\bar D}^{(*)} \Sigma^{(*)}_c$
molecule \cite{cpc,prl1}.

Through the S-wave charmed meson and baryon scattering, the
hidden-charm baryons with negative parity can also be generated
dynamically \cite{zou}. The total widths of the hidden-charm baryons
were less than 60 MeV, quite narrow. The charm-less decay modes are
important within this formalism.

The two $P_c$ states were also explained as the tightly bound states
\cite{polosa,prl2}. For example, the authors of Ref. \cite{polosa}
assumed quarks and diquarks are fundamental building blocks in the
diquark model. The mass difference between $P_c(4380)$ and
$P_c(4450)$ is about 70 MeV, which is - partly due to the orbital
excitation around 280 MeV and partly due to the mass difference
between the scalar and axial-vector diquarks around 200 MeV.

The other possible interpretations of these $P_c$ states can be
found in the review \cite{report}.

\subsection{X(3872)}

In 2003, Belle collaboration observed X(3872) in the $J/\psi \pi
\pi$ mode \cite{belle-x3872}, which was the first XYZ state. The
$J^{PC}$ quantum numbers of X(3872) are $1^{++}$. Its production
rate at the hadron colliders (Tetraron and LHC) is similar to that
of $\psi'$. The quark model prediction for the $\chi_{c1}'$ mass is
roughly 100 MeV higher, where the $\chi_{c1}'$ is the radial
excitation of the axial vector charmonium.

X(3872) lies very close to the $\bar D D^*$ threshold with a mass
difference less than 0.2 MeV. This state is very narrow. Its total
width is less than 1 MeV.  The discovery mode $X(3872)\to J/\psi
\rho \to J/\psi \pi \pi$ violates isospin symmetry, but its decay
width is comparable to the decay width of $X(3872)\to J/\psi \omega
\to J/\psi \pi \pi \pi$ decay mode. One may wonder whether X(3872)
is an axial-vector charmonium or molecular state.

Within the meson exchange model, we considered (1) the S-D wave
mixing which plays an important role in forming the loosely bound
deuteron; (2) the mass difference between the neutral and charged
$D/D^*$ mesons, and (3) the coupling of $\bar D D^*$ to ${\bar
D}^*D^*$ channel \cite{lining-x3872,m2,m3}. We notice that X(3872)
is a good candidate of the loosely bound molecular state. In fact,
if we replace the proton and neutron inside the deuteron by the
$\bar D$ and $D^*$ mesons, we reproduce the X(3872). Within the
molecular scheme, the large isospin violation can be explained
naturally \cite{lining-x3872}.

However, the E1 decay pattern suggests that X(3872) is a good
candidate of the axial vector charmonium
\cite{babar-gamma,lhcb-gamma}. If X(3872) is a radial excitation of
$\chi_{c1}$, both the radial wave functions of $\chi_{c1}'$ and
$\psi(2S)$ contain one node. Their overlapping is large.
$\chi_{c1}'$ will decay into $\psi(2S) \gamma$ more easily. In fact,
the experimental E1 decay rate of X(3872) is consistent with the
quark model prediction for the $\chi_{c1}'$.

Based on the measurement of the E1 decay ratio, LHCb concludes: "The
measured value agrees with expectations for a pure charmonium
interpretation of X(3872) and a molecular-charmonium mixture
interpretations" \cite{lhcb-gamma}. Moreover, the large production
cross section of X(3872) at LHC with very large $P_T$ is comparable
with that of $\psi(2S)$, which requires a significant $c\bar c$
component. On the other hand, the isospin violating dipion decay of
X(3872) requires the molecular component. The current experimental
information strongly suggests that X(3872) should probably be a
mixture of $\chi_{c1}'$ and $\bar D D^*$ molecule \cite{chao-2005}.

The recent dynamical lattice QCD simulation used many operators
including $c\bar c$, two-meson and diquark-antidiquark ones
\cite{lattice-x3872}. They found a lattice candidate for the X(3872)
with $J^{PC} = 1^{++}$ and I = 0 only if both the $c\bar c$ and
$\bar D D^*$ operators are included. This candidate cannot be found
without the $c\bar c$ component. This lattice QCD simulation
strongly supports X(3872) as a mixture of $c\bar c$ and molecule.

It's interesting to compare three candidates of the exotic states:
$\Lambda(1405)$, $D_{sj}(2317)$ and X(3872). $\Lambda(1405)$ is
lower than the quark model prediction for the P-wave $uds$ state and
lies very close to the $\bar K N$ threshold. $D_{sj}(2317)$ is lower
than the quark model prediction for the P-wave charm strange meson
and lies very close to the $DK$ threshold. X(3872) is lower than the
quark model prediction for the P-wave $c\bar c$ state $\chi_{c1}'$
and lies very close to the $\bar D D^*$ threshold.

In the above three cases, we observe the common feature: the couple
channel effects play a very important role and lower the bare quark
model level significantly. The S-wave continuum couples to the bare
quark model state strongly. The quark model spectrum is distorted
dramatically. For comparison, the bottomonium analogue $X_b$ was not
found since $\chi_{b1}'$ is not close to the $\bar B B^*$ threshold.

\subsection{The charged $Z_b$ and $Z_c$ states}

Let's move on to the charged states. In 2011, Belle collaboration
observed two charged $Z_b$ states \cite{bellezb}. They are very
close to the $\bar B B^*$ and ${\bar B}^* B^*$ threshold with
$J^P=1^+$. Their open-bottom decay modes are dominant. Later, BESIII
\cite{bes1} and Belle \cite{bellezc} collaborations observed a
similar state $Z_c(3900)$ in the $J/\psi \pi$ mode, which is close
to the $\bar DD^*$ threshold with $J^P=1^+$. This state was also
observed in the $\bar DD^*$ mode. Compared with the traditional
charmonium, the open-charm decay mode of $Z_c(3900)$ is strongly
suppressed. The decay dynamics might be different. These $Z_b$ and
$Z_c$ states are very similar.

One may wonder whether $Z_b$ and $Z_c$ are tetraquark states. If
they are tetraquarks, they shall fall apart into the open-charm
modes easily. The s-wave $\bar DD^*$ mode should dominate the ${\bar
D}^* D^*$ mode for the higher state $Z_c(4025)$ because of the huge
phase space difference. However, BESIII didn't observe $\bar DD^*$
mode for $Z_c(4025)$ up to now while Belle didn't observe $\bar B
B^*$ mode for the higher $Z_b$ state.

Let's turn back to the above dynamical lattice QCD simulation, which
used many operators. They didn't find any exotic candidates in the
isovector channel. This lattice QCD simulation strongly disfavors
either the diquark-antidiquark or tetraquark interpretations of the
X(3872) and $Z_c(3900)$ \cite{lattice-x3872}.

If the $Z_b$ states are real resonances, they could be the S-wave
${\bar B}^{(*)} B^*$ molecular states \cite{m1,m2,m3}. In fact,
within the meson exchange model, both $Z_b$ states can be explained
as the $\bar BB^*$ and ${\bar B}^*B^*$ molecules. Besides the
isovector $Z_b$ states, there are also several loosely bound
isoscalar molecular states. However, within the same model, the
$Z_c$ states seem unbound with a "reasonable" cutoff parameter
\cite{m4,m5}. The potential is roughly the same for the $Z_b$ and
$Z_c$ systems. But the kinetic energy of the $Z_c$ systems is larger
since the D meson mass is smaller than the B meson mass. The
hidden-charm/bottom molecules were discussed extensively in
literature \cite{m4,m5,v1,v2,hosaka}.

The $Z_c$ states lie above the open-charm thresholds. Their measured
mass and width seem channel dependent. Could they be non-resonant
signals arising from open-charm/bottom thresholds, final state
interactions such as $\bar DD^*$ rescattering or triangle
singularity etc? Some of these non-resonant mechanisms could explain
the current experimental data.

\subsection{Y(4260)}

In PDG, there are three well-established vector charmonium above 4
GeV: $3S/\psi(4040)$, $2D/\psi(4160)$, $4S/\psi(4415)$. In the quark
model, one expects at most five vector charmonium states between 4
and 4.7 GeV: $3S/\psi(4040)$, $2D/\psi(4160)$, $4S/\psi(4415)$,
$3D$, $5S$. But seven states were observed experimentally:
$\psi(4008)$, $\psi(4040)$, $\psi(4160)$, $\psi(4260)$,
$\psi(4360)$, $\psi(4415)$, $\psi(4660)$. What are these additional
Y states? The situation is very confusing now.

Y(4260) was first discovered in the $J/\psi \pi\pi$ mode with the
ISR technique by Babar collaboration \cite{babar-y4260} while
Y(4360) was observed in the $\psi(2S)\pi\pi$ channel with the same
technique \cite{babar-y4360}. But these two states were not observed
in the R-value scan and open-charm process. In the R-value scan, all
the well-established vector charmonium appear as a peak. But Y(4260)
and Y(4360) show up as a dip.

Y(4260) may be accommodated as the $\psi(4S)$ charmonium state with
the screened linear potential \cite{chao}. Y(4260) also seems a very
good candidate of the charmonium hybrid \cite{zhu-y4260,kou,page}.
According to lattice QCD simulation \cite{lattice-liu,lattice-chen},
the vector hybrid charmonium lie around 4.26 GeV. Because of the
gluon, the $1^{--}$ hybrid charmonium does not couple to the virtual
photon very strongly, which explains the dip in the R-value scan.
One of the favorable decay mode of hybrid states is the p-wave +
s-wave meson pair, which explains the non-observation in the
$D^{(*)}{\bar D}^{(*)}$ modes. The $c\bar c$ pair within the vector
charmonium is a spin-singlet while the gluon is color-magnetic,
which is favorable to the spin-singlet hidden-charm decay mode.

\section{Summary}

Now let me summarize. The excited Upsilon states act as a molecule
factory. Because $M[\Upsilon(5S)]=10.860$ GeV,
$M[BB^*+\pi]=10.604+0.140=10.744$ GeV, and $
M[B^*B^*+\pi]=10.650+0.140=10.790$ GeV, the phase space of the decay
$\Upsilon(5S)\to {\bar B}^{(*)}B^* \pi$ is tiny. The relative motion
between the ${\bar B}^{(*)}B^*$ pair is very slow, which is
favorable to the formation of the ${\bar B}^{(*)}B^*$ molecular
states. $\Upsilon(5S)$ or $\Upsilon(6S)$ is the ideal place to study
either the molecular states or the ${\bar B}^{(*)}B^*$ interaction.
Similar signals will be produced abundantly at Belle2 in the coming
years!

The vector charmonium spectrum is very puzzling at present. The
excited charmonium decay is ideal in the search of the $Z_c$
signals. The $\gamma$, $1\pi$, $2\pi$, $3\pi$ and other light degree
of freedom will act as a quantum number filter of these states.
X(3872), $\chi_{c1}'$, and Y(4260) are the key states in revealing
the underlying structure of the charmonium-like XYZ states. Through
the decay pattern and possible partner states, we can test the
various theoretical picture. The experimental measurement of the
various pionic and electromagnetic transitions are crucial.

\section*{Acknowledgments}

The authors are grateful to X.L. Chen, M.L. Du, W.Z. Deng, N. Li,
X.H. Liu, Y.R. Liu, Z.G. Luo, L. Ma, T.G. Steele, Z.F. Sun, and L.
Zhao for fruitful collaborations. This project was supported by the
973 program and National Natural Science Foundation of China under
Grants 11575008 and 11621131001.

\end{document}